\documentclass[reprint,aps,prb,superscriptaddress,twocolumn,floatfix,amsmath,amssymb,amsfonts,longbibliography]{revtex4-2}
\usepackage{graphicx}% Include figure files
\usepackage[
colorlinks, 
citecolor=blue, 
urlcolor=blue, 
linkcolor=red,breaklinks]{hyperref}
\usepackage[utf8]{inputenc}
\usepackage[vietnamese, english]{babel}
\usepackage{braket}
\usepackage{xcolor}
\usepackage{soul}

\begin{document}

\title{Chern non-trivial two-dimensional square-net nodal line semimetal}

\author{Kendall Ramos}
\affiliation{Department of Physics, University of Arkansas, Fayetteville, AR 72701, USA}

\author{Salvador Barraza-Lopez}
\email{sbarraza@uark.edu}
\affiliation{Department of Physics, University of Arkansas, Fayetteville, AR 72701, USA}

\author{Elias Andrade}
\email{eandrade@uark.edu}
\affiliation{Department of Physics, University of Arkansas, Fayetteville, AR 72701, USA}

\date{\today}

\begin{abstract}
We investigate a two-dimensional $p_xp_y$ square-net tight-binding model with two sublattices and introduce a symmetry-informed, momentum-dependent spin-orbit coupling (SOC) with a finite-range real-space hopping representation. Its two components jointly gap the nodal line, while one breaks time-reversal symmetry, producing a Chern-insulating phase with an occupied band Chern number $C=2$. Nanoribbon calculations reveal edge-localized states traversing this gap, with two net chiral modes per edge. We further show that the conserved spin sectors are unitarily equivalent and therefore have equal Chern numbers, ruling out a nontrivial time-reversal-protected $\mathbb{Z}_2$ phase within this ansatz. Our results establish a phenomenological route to Chern topology in square-net two-dimensional systems and clarify the symmetry constraints governing the accessible insulating phases.
\end{abstract}

\maketitle

\section{Introduction}\label{sec:intro}

Topological semimetals exhibit band crossings whose stability is governed by symmetry and topology. Whereas nodal point semimetals host isolated crossings, nodal line semimetals feature one-dimensional manifolds of band degeneracies in momentum space~\cite{armitage_weyl_2018,fang_topological_2016,yu_topological_2017}. Perturbations that lift these degeneracies provide a route to insulating phases, whose topology depends on both the structure of the gap-opening coupling and the symmetries of the carrier dynamics that remain.

ZrSiS was first shown to host a
nonsymmorphic-symmetry-protected three-dimensional Dirac line node by
Schoop \emph{et al.}\ \cite{schoop_dirac_2015}. Angle-resolved
photoemission spectroscopy subsequently confirmed both nodal surfaces
and nodal lines \cite{fu_dirac_2019} and established a topological
nodal fermion semimetal phase \cite{neupane_observation_2016} in
which bands cross along extended lines in the first Brillouin zone rather
than at isolated points, producing a network of nodal lines close to
the Fermi level, illustrating the rich electronic structure associated with these systems~\cite{chen_dirac_2017,singh_saddlelike}. Although ZrSiS is a three-dimensional material, two-dimensional $p_xp_y$ square-net models capture important features of the bands associated with its Si layers~\cite{Klemenz2019topological,Klemnz2020role,Klemenz2020systematic}. These models also provide a tractable framework for studying how spin-dependent couplings modify nodal crossings to generate topological gaps.

The broader interest in two-dimensional nodal line systems extends to proposed carbon-nitride networks, strained black-phosphorus films, octastanene halogenides, magnetic monolayers and metal--organic frameworks~\cite{chen_prediction_2018,liu_prediction_2020,zhang_octastanene_2023,jin_two-dimensional_2020,wang_half-metals_2022}. Experimental evidence for two-dimensional Dirac nodal line fermions has also been reported in Cu$_2$Si monolayers~\cite{feng_experimental_2017}, following theoretical predictions of its planar structure~\cite{yang_two-dimensional_2015}. Related studies have explored edge-state phenomena, SOC-induced modifications of nodal lines, and the engineering of type-II nodal line dispersions~\cite{dai_two-dimensional_2024,fernandez_two-dimensional_2024,chen_engineering_2026}, as well as collective excitations and screening~\cite{rahimpoor_collective_2024}. A broader overview is provided in Ref.~\cite{Wadge_2026}.

Previous studies have investigated gap opening in square-net models through SOC combined with sublattice asymmetry, introduced through lattice distortions, buckling, or additional sublattice-dependent terms~\cite{luo2015room,deng2022twisted,aryal2022topological}. More recently, a sublattice-asymmetric effective SOC was obtained by projecting a larger orbital model onto the low-energy states of a $p_xp_y$ model inducing a band gap~\cite{orozco-galvan_tight-binding_2024}. However, opening a gap does not by itself establish a nontrivial insulating phase. In particular, having a nonzero Chern number over all occupied bands requires broken time-reversal symmetry~\cite{haldane_model_1988}, whereas conventional SOC in a nonmagnetic system preserves this symmetry.

In this work, we retain the matrix structure of the effective coupling proposed in Ref.~\cite{orozco-galvan_tight-binding_2024} and engineer its momentum dependence over the full Brillouin zone. We show that the resulting Hamiltonian separates into two conserved spin sectors related by a momentum-independent unitary transformation. The equal Chern numbers at these two spin sectors exclude a nontrivial time-reversal-protected $\mathbb{Z}_2$ phase within this ansatz~\cite{kane_z2_2005,Kane2005quantum}, while allowing the two sectors to contribute equally to a Chern-insulating phase when time reversal is broken. Guided by this constraint, we introduce a finite-range hopping pattern whose two components jointly open a bulk gap with an occupied band Chern number $C=2$. Nanoribbon calculations reveal edge-localized branches traversing the gap, with two net chiral modes per edge. Our construction provides a phenomenological framework for engineering Chern topology in square-net systems without assuming a microscopic realization of the proposed coupling in ZrSiS.

The paper is organized as follows: In Sec.~\ref{sec:model} we introduce the square-net tight-binding model and the spin-dependent coupling. In Sec.~\ref{sec:topology} we analyze the symmetry constraints, present the momentum-dependent ansatz, and evaluate the bulk topology. In Sec.~\ref{sec:edges} we examine the nanoribbon spectrum and its relation to the bulk Chern number. Findings are summarized in Sec.~\ref{sec:Conclusion}.

\section{Model}\label{sec:model}
\begin{figure}[t]
\begin{center}
{\includegraphics[scale = 1]{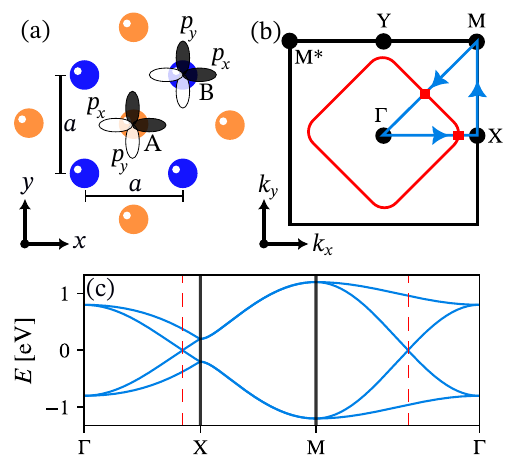}}
\caption{(a) Square lattice with two atoms per unit-cell, the sublattices are labeled as A and B and indicated by orange (light) and blue (dark) circles, respectively. Each of the atoms contributes to the tight-binding Hamiltonian with $p_x$ and $p_y$ orbitals, shown as well. (b) First Brillouin zone of the two-atom square lattice. The black circles indicate the high-symmetry points. The spectrum features a nodal line indicated schematically by the (red) curve. The path-shown is used to plot the dispersion in (c); squares show the points where the path crosses the nodal line.}
\label{fig:Lattice}
\end{center}
\end{figure}

The nodal line featured in ZrSiS can be described by the $p$-orbitals in the Si planes, forming a square-net as shown in Fig.~\ref{fig:Lattice}(a)~\cite{orozco-galvan_tight-binding_2024}. The inclusion of two atoms per unit cell results in the folding of the first Brillouin zone. For square-net $p_xp_y$ models, this results in the appearance of a nodal line, where valence and conduction bands become degenerate along a curve in momentum space, forming a nodal line semimetal. The path of the nodal line within the first Brillouin zone is shown schematically in Fig.~\ref{fig:Lattice}(b). The blue arrows indicate the $\boldsymbol{k}$-path used to plot the spectrum in Fig.~\ref{fig:Lattice}(c). The vertical dashed lines on Fig.~\ref{fig:Lattice}(c) indicate the crossing of the nodal line, which is linear.  

Previous theoretical works have investigated SOC in models relevant to ZrSiS by introducing sublattice asymmetry through several mechanisms, including a staggered potential~\cite{aryal2022topological}, in-plane lattice distortions~\cite{deng2022twisted}, lattice buckling~\cite{luo2015room}, and sublattice-asymmetric SOC~\cite{orozco-galvan_tight-binding_2024}. Here, we build on the last approach, which retains Si atoms in a two-dimensional square-net and uses L\"owdin partitioning~\cite{lowdin_studies_1963} to project the SOC contribution of a larger orbital model onto an effective Hamiltonian (dubbed $P_{\text{SOC}}$) for the low-energy states.

 Prior to the inclusion of the SOC term, the $p_xp_y$ tight-binding Hamiltonian for the square-net with two atoms per unit cell is:  
\begin{equation}
H_0(\boldsymbol{k}) =
\begin{pmatrix}
\alpha(\boldsymbol{k})_x & 0 & \beta(\boldsymbol{k}) & \gamma(\boldsymbol{k}) \\
0 & \alpha_y(\boldsymbol{k}) & \gamma(\boldsymbol{k}) & \beta(\boldsymbol{k}) \\
\beta(\boldsymbol{k}) & \gamma(\boldsymbol{k}) & \alpha_x(\boldsymbol{k}) & 0 \\
\gamma(\boldsymbol{k}) & \beta(\boldsymbol{k}) & 0 & \alpha_y(\boldsymbol{k})
\end{pmatrix},
\label{eq:Hfold}
\end{equation}
with:
\begin{subequations}
\begin{equation}
\alpha_x(\boldsymbol{k}) = 2t_{\sigma_2}\cos(k_xa) + 2t_{\pi_2}\cos(k_ya),
\end{equation}

\begin{equation}
    \alpha_y(\boldsymbol{k}) = 2t_{\sigma_2}\cos(k_ya) + 2t_{\pi_2}\cos(k_xa),
\end{equation}
\begin{equation}
        \beta(\boldsymbol{k}) = (t_{\sigma_1} +t_{\pi_1})\left[\cos\left(\frac{k_x+k_y}{2}a \right) + \cos\left(\frac{k_x-k_y}{2}a \right)\right],
\end{equation}
\begin{equation}
    \gamma(\boldsymbol{k}) = (t_{\sigma_1} - t_{\pi_1})\left[\cos\left(\frac{k_x+k_y}{2}a \right) - \cos\left(\frac{k_x-k_y}{2}a \right)\right],
\end{equation}
\end{subequations}
where $t_{\sigma_1}=500$ meV ($t_{\sigma_2}=50$ meV) is the first (second) nearest neighbor hopping parameter for $\sigma$-bonds and similarly, $t_{\pi_1}=-100$ meV ($t_{\pi_2}=-50$ meV) is the first (second) nearest neighbor hopping for $\pi$-bonds~\cite{Klemenz2020systematic}. The Hamiltonian is written on the \{$|p^A_{x}\rangle, |p^A_{y}\rangle, |p^B_{x}\rangle, |p^B_{y}\rangle\}$ basis.

The Hamiltonian can be compactly written in terms of outer products:
\begin{equation}
\begin{split}
H_0(\boldsymbol{k})=&d_0(\boldsymbol{k})\tau_0\otimes\upsilon_0+d_z(\boldsymbol{k})\tau_z\otimes\upsilon_0\\
&+\beta(\boldsymbol{k})\tau_0\otimes\upsilon_x+\gamma(\boldsymbol{k})\tau_x \otimes \upsilon_x,
\label{Eq:H0}
\end{split}
\end{equation}
where:
\begin{subequations}
    \begin{equation}
        d_0(\boldsymbol{k})=\frac{\alpha_x(\boldsymbol{k})+\alpha_y(\boldsymbol{k})}{2},
    \end{equation}
    \begin{equation}
        d_z(\boldsymbol{k})=\frac{\alpha_x(\boldsymbol{k})-\alpha_y(\boldsymbol{k})}{2},
    \end{equation}
\end{subequations}
and \{$\tau_0$, $\tau_x$, $\tau_y$, $\tau_z$\} and \{$\upsilon_0$, $\upsilon_x$, $\upsilon_y$, $\upsilon_z$\} are Pauli matrices acting on the orbital and sublattice degrees of freedom, respectively. The electronic spectrum is found to be:
\begin{subequations}
\begin{equation}
    E_{e,\pm}(\boldsymbol{k})=d_0(\boldsymbol{k})+|\beta(\boldsymbol{k})\pm \epsilon(\boldsymbol{k})|,
\end{equation}
\begin{equation}
    E_{h,\pm}(\boldsymbol{k})=d_0(\boldsymbol{k})-|\beta(\boldsymbol{k})\pm \epsilon(\boldsymbol{k})|,
\end{equation}
\end{subequations}
with:
\begin{equation}
\epsilon(\boldsymbol{k})=\sqrt{d_z^2(\boldsymbol{k})+\gamma^2(\boldsymbol{k})}.
\end{equation}
The condition for the existence of a nodal line is then $\beta^2(\boldsymbol{k})=d_z^2(\boldsymbol{k})+\gamma^2(\boldsymbol{k})$.

The Hamiltonian containing the SOC term is:
\begin{equation}
    H(\boldsymbol{k})=H_0(\boldsymbol{k})\otimes\sigma_0 + P_{\text{SOC}},
\end{equation}
where following Ref.~\cite{orozco-galvan_tight-binding_2024}, the coupling has the matrix structure:
\begin{equation}
\label{Eq:Psoc}
\begin{split}
    P_{\text{SOC}}=-\mathbb{R}\text{e}[\eta(\boldsymbol{k})]\tau_x \otimes\upsilon_z\otimes \sigma_x + \mathbb{I}\text{m}[\eta(\boldsymbol{k})]\tau_x \otimes\upsilon_y\otimes \sigma_x,\\
    %&&=Q\otimes\sigma_x.
    \end{split}
\end{equation}
where \{$\sigma_0$, $\sigma_x$, $\sigma_y$, $\sigma_z$\} are Pauli matrices acting on the spin degree of freedom. $P_{\text{SOC}}$ lifts the degeneracy along the nodal line, as demonstrated in Ref.~\cite{orozco-galvan_tight-binding_2024}. There, the matrix structure was obtained through a projection near a band crossing, where $\eta$ was approximated by a constant complex number $\eta\neq\eta(\boldsymbol{k})$. In the following section, we retain the matrix structure and propose an explicit momentum dependence over the first Brillouin zone to turn the Chern invariant nontrivial.

\section{Inducing a Non-Trivial Topology}\label{sec:topology}

Since $P_{\text{SOC}}$ opens a gap, we investigate the topology of the resulting insulating phases. For the present $P_{\text{SOC}}$, however, this coupling cannot realize a nontrivial $\mathbb{Z}_2$ insulator. Since $P_{\text{SOC}}=Q\otimes\sigma_x$, $[H_0\otimes\sigma_0+P_{\text{SOC}},\sigma_x]=0$, the Hamiltonian separates into two conserved spin sectors,  $H_{\pm}=H_0\pm Q$. These sectors are related by
\begin{equation}
    H_-=VH_+V^\dagger, \quad V=\tau_0\otimes\upsilon_x.
\end{equation}
Because $V$ is momentum independent, the occupied bands of both sectors have equal Chern numbers, $C_+=C_-$. If time-reversal symmetry is imposed, it additionally requires $C_+=-C_-$, forcing both sector Chern numbers to vanish and ruling out a quantum spin Hall insulator.

\begin{figure}[t]
\begin{center}
{\includegraphics[scale = 1]{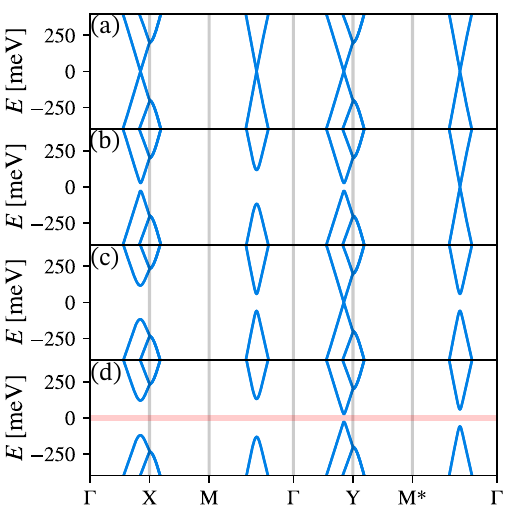}}
\caption{Band structures with the introduction of $P_{\text{SOC}}$ with $\eta(\boldsymbol{k})$ as defined in Eq.~(\ref{Eq:eta(k)}) for different values of $\mu$ and $\lambda$. (a) $\mu=\lambda=0$. (b) $\mu=30$ meV and $\lambda=0$. (c) $\mu=0$ and $\lambda=30$ meV. (d) $\mu=\lambda=30$ meV, this is the only case where a gap is opened, as highlighted by the shaded region.}
\label{fig:Bands}
\end{center}
\end{figure}

We therefore turn to couplings $\eta(\boldsymbol{k})$ that break time-reversal symmetry and permit insulating phases with a nonzero Chern number. This is, we seek functions $\eta(\boldsymbol{k}$) satisfying:
\begin{equation}
\mathcal{T}P_{\text{SOC}}(\boldsymbol{k})\mathcal{T}^{-1}
\neq P_{\text{SOC}}(-\boldsymbol{k}),
\end{equation}
where $\mathcal{T}=(\tau_0 \otimes\upsilon_0\otimes i\sigma_y)K$ is the time-reversal operator, with $K$ denoting complex conjugation. This leads to the condition that either:
\begin{equation}
    \mathbb{R}\text{e}[\eta(-\boldsymbol{k})]\neq -\mathbb{R}\text{e}[\eta(\boldsymbol{k})], \quad 
\end{equation}
or: 
\begin{equation}
    \mathbb{I}\text{m}[\eta(-\boldsymbol{k})]\neq \mathbb{I}\text{m}[\eta(\boldsymbol{k})].
\end{equation}
The real-part contribution to $P_{\text{SOC}}$ in Eq.~(\ref{Eq:Psoc}) is diagonal in sublattice space through $\upsilon_z$, and given that it describes coupling within each sublattice, we represent it through a second nearest neighbor interaction. The imaginary-part contribution to $P_{\text{SOC}}$ in Eq.~(\ref{Eq:Psoc}) is off-diagonal in sublattice space through $\upsilon_y$, and we represent it using nearest-neighbor interaction between the two sublattices. Specifically, we choose:
\begin{equation}
\begin{split}
\eta(\boldsymbol{k})=&2\mu[\sin(k_xa)+\sin(k_ya)]\\
&+4i\lambda\left[\sin\left(\frac{k_xa}{2}\right)\cos\left(\frac{k_ya}{2}\right)\right],
\end{split}
\label{Eq:eta(k)}
\end{equation}
where $\mu$ and $\lambda$ are real parameters. The real part is odd under momentum reversal and satisfies the time-reversal condition, whereas the imaginary part is also odd and therefore breaks time-reversal symmetry when $\lambda\neq0$. The bands are 2-fold degenerate since the spin subsectors $H_{\pm}$ have the same spectrum, given by,
\begin{subequations}
\begin{equation}
    E^\eta_{e,\pm}(\boldsymbol{k})=d_0(\boldsymbol{k})+\sqrt{|\beta(\boldsymbol{k})\pm \epsilon(\boldsymbol{k})|^2+|\eta(\boldsymbol{k})|^2},
\end{equation}
\begin{equation}
    E^\eta_{h,\pm}(\boldsymbol{k})=d_0(\boldsymbol{k})-\sqrt{|\beta(\boldsymbol{k})\pm \epsilon(\boldsymbol{k})|^2+|\eta(\boldsymbol{k})|^2}.
\end{equation}
\end{subequations}
The electronic bands are shown in Fig.~\ref{fig:Bands} for different parameters of $\mu$ and $\lambda$. The inclusion of a finite value for only $\mu$ or $\lambda$ lifts the degeneracy at some points of the nodal line. However, they alone are not enough to open a gap through the entire first Brillouin zone, as shown in Figs.~\ref{fig:Bands}(b) and (c), respectively. Only when $\mu\neq0$ and $\lambda\neq0$ a gap is opened, as shown in Fig~\ref{fig:Bands}(d). For $\mu=\lambda=30$ meV the resulting gap has a magnitude $\sim 37$ meV, as highlighted by the shaded region on Fig~\ref{fig:Bands}(d). 

To calculate the Chern number of the occupied bands, we use the lattice-gauge-invariant method of Fukui, Hatsugai, and
Suzuki~\cite{fukui_chern_2005}. When a group of $N$ bands is degenerate, or bands touch, individual band Chern numbers are no longer
meaningful. However, the total Chern number of the full $N$-dimensional
subspace can still be computed through a non-Abelian generalization. Defining:
\begin{equation}
U_{\boldsymbol{q}}(\boldsymbol{k}) =
\frac{\det [\Psi^{\dagger}(\boldsymbol{k})\
\Psi(\boldsymbol{k}+
\boldsymbol{q})]}
{\big|\det [\Psi^{\dagger}(\boldsymbol{k})\
\Psi(\boldsymbol{k}+\boldsymbol{q})]\big|},
\label{eq:linkvar_multi}
\end{equation}
where $\Psi(\boldsymbol{k})$ is a matrix whose columns contain the wavefunctions of the occupied bands at momentum $\boldsymbol{k}$ and $\boldsymbol{q}$ is a finite excursion in reciprocal space. This non-Abelian construction does not produce Chern numbers for individual bands making up the subspace; it returns only the gauge-invariant total. The lattice field strength at a given $\boldsymbol{k}$-point is:
\begin{equation}
F(\mathbf{k}) = \ln \big[U_{\boldsymbol{q}_x}(\boldsymbol{k})\,
U_{\boldsymbol{q}_y}(\boldsymbol{k}+\boldsymbol{q}_x)\,U_{\boldsymbol{q}_x}(\boldsymbol{k}+{\boldsymbol{q}_y})^{-1}\,
U_{\boldsymbol{q}_y}(\boldsymbol{k})^{-1}\,\big],
\label{eq:fieldstrength}
\end{equation}
and the total Chern number contributed by the occupied bands is,
\begin{equation}
    C=\frac{1}{2\pi i} \sum_{\boldsymbol{k}\in \text{BZ}} F(\boldsymbol{k}),
\end{equation}
(the sum was performed in a grid of $600 \times600$ $\boldsymbol{k}$-points within the first Brillouin zone.) For the lattice with $\eta(\boldsymbol{k})$ given in Eq.~(\ref{Eq:eta(k)}), we calculate the Chern number using the subspace spanned by the two lowest-energy bands of each spin-subsector Hamiltonian, $H_{\pm}$. We obtain $C_+=C_-=1$, yielding a total Chern number of $C=2$. Alternatively, calculating the Chern number using the subspace spanned by the four occupied lowest-energy bands of the full Hamiltonian $H$ yields the same result, $C=2$.

\section{Bulk-Boundary Correspondence}\label{sec:edges}
\begin{figure}[hbtp]
\begin{center}
{\includegraphics[scale = 1]{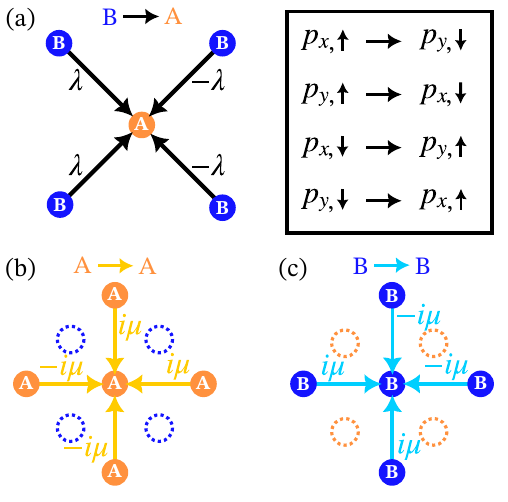}}
\caption{Mapping of the SOC $\eta(\boldsymbol{k})$ in real space. (a) Interlattice interaction at first nearest neighbors, the magnitude of the interaction $\lambda$ changes sign while moving along the $x$-direction. (b)-(c) Intralattice interaction for second nearest-neighbors with a imaginary amplitude $\mu$. The signs are opposite between sublattices. In each process there is a flip in the orbital and spin as indicated in the top-right inset.}
\label{fig:SpaceEta}
\end{center}
\end{figure}
Next we look into the topological edge states arising from the nonzero Chern number. First, we map $P_{\text{SOC}}(\boldsymbol{k}) $ into real space; this process is schematically depicted as shown on Fig.~\ref{fig:SpaceEta}. As discussed in Sec.~\ref{sec:topology}, there are two contributions to $\eta(\boldsymbol{k})$: the imaginary part introduces a first-nearest-neighbor interaction of strength $\lambda$ that changes sign between neighbors along the $x$-direction [Fig.~\ref{fig:SpaceEta}(a)]. On the other hand, the real part introduces a second-nearest-neighbor interaction $i\mu$ which changes sign depending on bond orientation, hopping direction, and sublattice [Figs.~\ref{fig:SpaceEta}(b)-(c)]. Both contributions couple states with opposite orbital and spin labels in the chosen basis.
\begin{figure*}[t!]
\begin{center}
{\includegraphics[scale = 1]{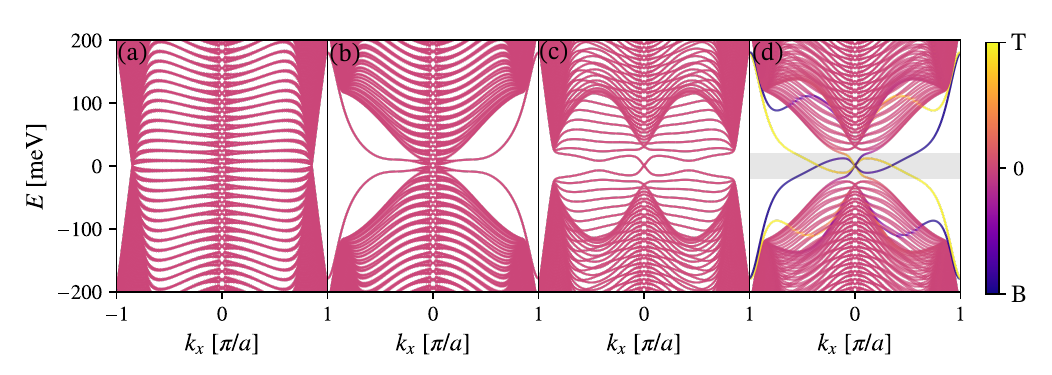}}
\caption{Band structure for nanoribbons of width $100a$. (a) $\mu=0$ and $\lambda=0$, (b) $\mu=30$ meV and $\lambda=0$, (c) $\mu=0$ and $\lambda=30$ meV and (d) $\mu=30$ meV and $\lambda=30$ meV. The bands are colored according to their localization in the transverse direction of the ribbon, such that T and B stand for top and bottom edge, respectively. Shading in (d) indicates the bulk gap.}
\label{fig:EdgeStates}
\end{center}
\end{figure*}

We are now able to calculate the electronic bands for nanoribbons, and we consider a periodic structure along the $x-$axis with a width $W=100a$ along the $y$-direction. The bands are shown in Fig.~\ref{fig:EdgeStates} for different values of $\mu$ and $\lambda$, and colored according to the expectation value $\braket{y}$to topologiclly tell edge states appart. When $\mu=\lambda=0$ [Fig.~\ref{fig:EdgeStates}(a)] there is no SOC and the spectrum is composed of ‘flat' bands with oscillations; these oscillations are a consequence of the nodal line in the bulk-spectrum, as multiple states at the same energy collapse in the one-dimensional spectrum and their avoided crossings results in oscillations. The cases for $\mu\neq0$, $\lambda=0$ and $\mu=0$, $\lambda\neq0$ are shown in Figs.~\ref{fig:EdgeStates}(b) and (c), respectively. Although each contribution modifies the band structure, neither opens a full bulk gap for the parameters considered [Figs.~\ref{fig:Bands}(b) and (c)] and the ribbon does not show edge states. Finally, when both $\lambda$ and $\mu$ are nonzero, a bulk gap opens [Fig.~\ref{fig:Bands}(d)], and edge-localized branches traverse this gap [Fig.~\ref{fig:EdgeStates}(d)]. The presence of two net chiral modes per edge, with opposite net propagation directions on opposite edges, is consistent with the bulk Chern number $C=2$.

\section{Conclusions}\label{sec:Conclusion}
We deployed a route to Chern-insulating phases in a two-dimensional $p_xp_y$ square-net semimetal through a momentum-dependent and spin-dependent coupling. Starting from a nodal line semimetal, we introduced a hopping pattern coupling the orbital, sublattice, and spin degrees of freedom and embed it with a $\boldsymbol{k}-$point dependence that allows for a simple real-space expression. The matrix structure constrains the accessible topological phases. Conservation of the spin component along $x$ separates the Hamiltonian into two unitarily equivalent sectors with equal Chern numbers. Imposing time-reversal symmetry forces both sector Chern numbers to vanish, excluding a nontrivial quantum spin Hall phase. Breaking time reversal symmetry instead allows the two sectors to contribute equally to a nonzero total Chern number.

For the parameters considered, its two components act together to open a full bulk gap, while the contribution proportional to $\lambda$ breaks time-reversal symmetry. For $\mu\neq0$ and $\lambda\neq0$ the occupied band subspace has Chern number $C=2$. Nanoribbon calculations exhibit edge-localized branches traversing the bulk gap, with two net chiral modes per edge, consistent with this bulk invariant.

 These results establish a phenomenological framework for engineering Chern topology in two-dimensional square-net models and identify the symmetry constraints that distinguish this construction from time-reversal-preserving SOC.
\section{Acknowledgements}
This work was supported by the US Department of Energy (Grant DE-SC0022120). We acknowledge useful conversations with G. Orozco-Galvan and J. Mella-Riquelme. 

\bibliography{references}

\end{document}